\documentclass[mathleft]{an}
\usepackage{graphicx}
\usepackage{times}
\begin{document}
\newcommand{\kms}{\,km\,s$^{-1}$}
\newcommand{\msec}{\,m\,s$^{-1}$}
\Pagespan{789}{}
\Yearpublication{2007}%
\Yearsubmission{2007}%
\Month{11}%
\Volume{999}%
\Issue{88}%

\title{Surface velocity network with anti-solar differential rotation on the active K-giant $\sigma$\,Geminorum
}

\author{Zs. K\H{o}v\'ari \inst{1}\fnmsep\thanks{Corresponding author:
  \email{kovari@konkoly.hu}\newline}
\and J.~Bartus\inst{2}
\and M.~\v{S}vanda\inst{3,4}
\and K.~Vida\inst{1,5}
\and K.~G.~Strassmeier\inst{2}
\and K.~Ol\'ah\inst{1}
\and E.~Forg\'acs-Dajka\inst{5}
}


\institute{ Konkoly Observatory, H-1525 Budapest, P.O.Box 67,
Hungary 
\and Astrophysical Institute Potsdam, An der Sternwarte
16, D-14482 Potsdam, Germany 
\and Astronomical Institute, Faculty of Mathematics and
Physics, Charles University in Prague, V Hole\v{s}ovi\v{c}k\'{a}ch
2, 180 00 Prague~8, Czech Republic
\and Astronomical Institute, Academy of Sciences of the Czech
Republic, v.~v.~i., Fri\v{c}ova 298, 251 65 Ond\v{r}ejov, Czech
Republic
\and E\"otv\"os University,
Department of Astronomy, H-1518 Budapest, P.O.Box. 32, Hungary}

\received{2007} \accepted{2007} \publonline{later}


\abstract{We demonstrate the power of the local correlation tracking technique on
stellar data for the first time. We recover the spot migration pattern of the
long-period RS\,CVn-type binary $\sigma$\,Gem from a set of six Doppler
images from 3.6 consecutive rotation cycles. The resulting surface flow map suggests a weak anti-solar differential rotation
with $\alpha\approx-0.0022\pm0.0016$, and
a coherent poleward spot migration with an average velocity of $220\pm10$\msec.
This result agrees with our recent findings from another study and could also
be confirmed theoretically.}

\maketitle

\section{Time-series Doppler images of $\sigma$\,Gem}
\label{sgemdi}

$\sigma$ Geminorum (75\,Gem, HR\,2973, HD\,62044) is a long-period
($P_{\rm rot}\approx 19.6$ days) RS\,CVn-type system with a K1-giant
primary and an unseen companion.
In our
initial paper (K\H{o}v\'ari et al. \cite{AAsgem})
six time-series Doppler images were recovered from the
Ca\,{\sc i}-6439\AA\ and Fe\,{\sc i}-6430\AA\ lines.
The maps cover 3.6 consecutive stellar rotations in 1996/97. Images \#1, \#3, and \#5, as
well as \#2, \#4, and \#6 represent contiguous stellar rotations and therefore independent maps
(see our extensive study for details).
Since there is a reasonable good agreement between the respective Ca and Fe maps,
for further use we combine them to decrease statistically the uncertainties
of the individual Doppler reconstructions.

\begin{figure*}
\includegraphics[angle=0,width=0.88\textwidth]{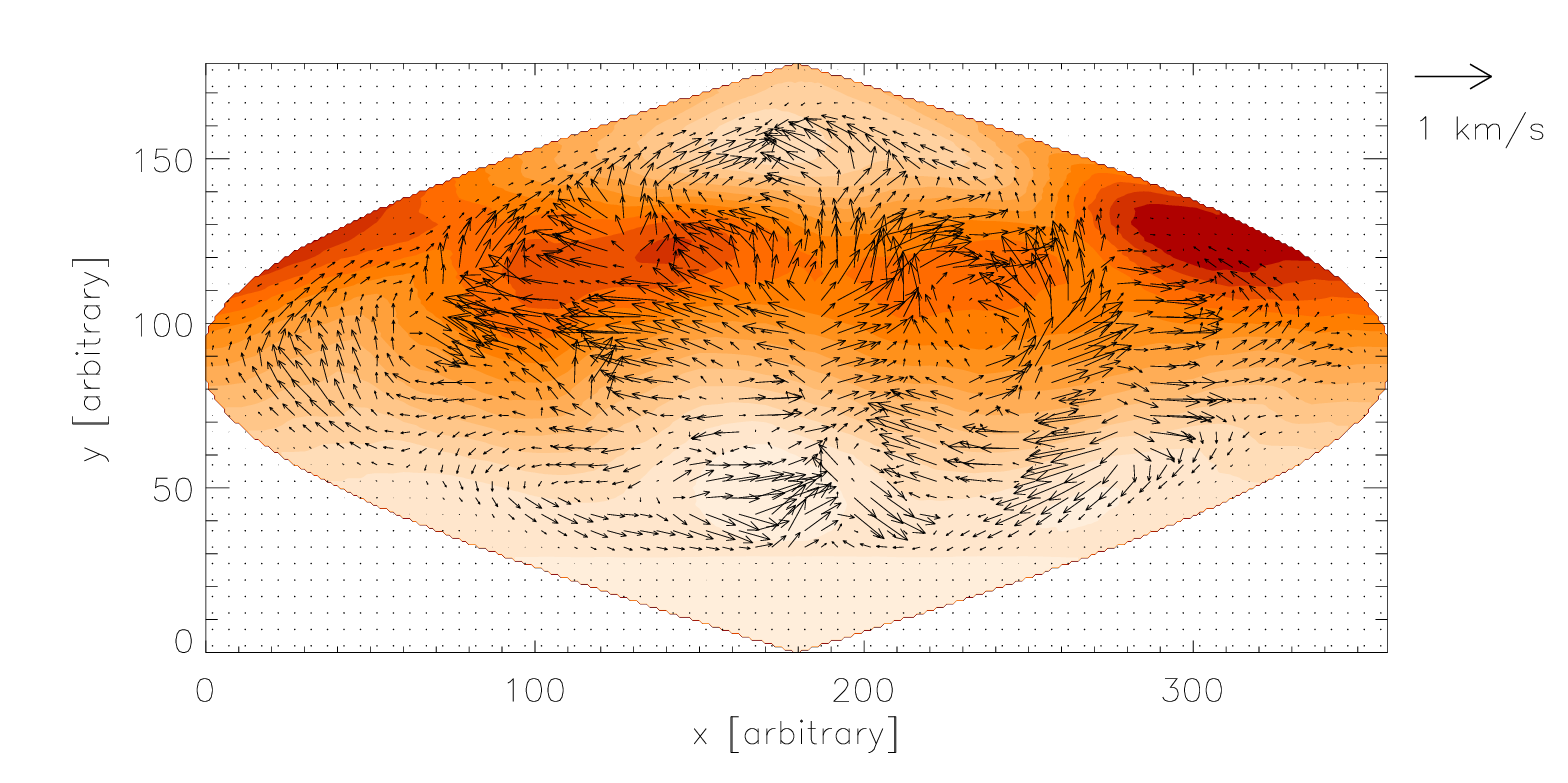}\hspace{-0.95cm}
\includegraphics[angle=0,height=7.4cm]{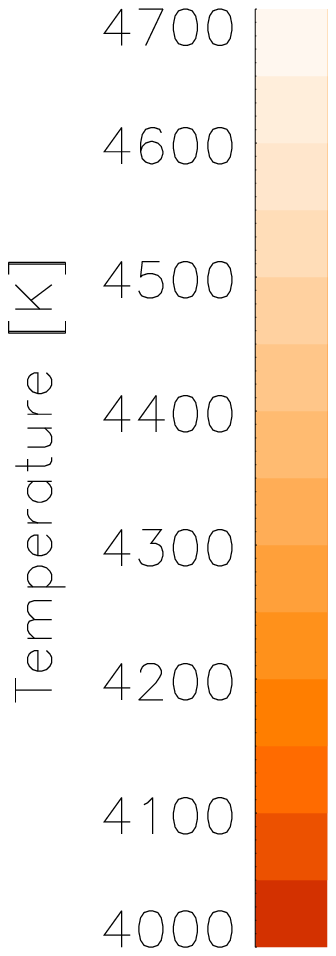}
\caption{Surface changes in the photosphere of $\sigma$\,Gem from LCT. As a background we
use the average of the six combined (Ca+Fe) time-series Doppler images.
Two distinct flow structures can be observed: a complex
network of convergent flows around spots and a general poleward trend.}
\label{flows}
\end{figure*}

\section{Local correlation tracking}\label{lct}

The local correlation tracking (hereafter LCT) technique, originally
developed for tracking different solar
surface features, cf., e.g., Sobotka et al. (\cite{sobo99},
\cite{sobo00}), Ambro\v{z} (\cite{amb}),
\v{S}vanda et al. (\cite{sva2}),
is based on the
principle of the best match of two image frames that record the
tracked surface structures at two different instants.
A small correlation window is chosen in the
first frame and compared with a somewhat displaced window in the
second frame. When the best matching displacement is found, a
vector of displacement is recorded, and a new correlation window
is chosen by the side of the former one. This pattern is followed
until the whole image frame is covered. The resulting vector field
is interpreted as the surface flow map over the time lag of the
two initial images. For a more detailed description of the method
as well as numerical simulations with artificial data, we refer to
a forthcoming paper (\v{S}vanda et al. \cite{lctlq}).

In our pilot study we apply the LCT technique to the
time-series of the six average (Ca+Fe) Doppler images of
$\sigma$\,Gem (K\H{o}v\'ari et al. \cite{AAsgem}).
From the six maps we set four
correlation pairs: \#1$-$\#3, \#2$-$\#4, \#3$-$\#5 and \#4$-$\#6, i.e., we
choose the independent consecutive but contiguous maps
to minimize the time shift and, thus, the masking effect of individual
spot evolution on the flow pattern. Then, for
each pair a velocity field map is computed and finally the four
maps are averaged.

The resulting average flow map is shown in Fig.~\ref{flows}. In
the figure the large-scale flow network is
related to the spatially resolved surface structures,
which is expected from solar analogy. The amplitude of the
velocity vectors of typically several hundreds of \msec\ also seem
reasonable. Flows of the order of 100\msec\ are observed in the
photosphere of the Sun.
Fig.~\ref{flows} shows convergent
flows towards spotted regions, which is also consistent with
the solar case. 

\begin{figure}[t]
 \includegraphics[angle=-90,width=0.97\columnwidth]{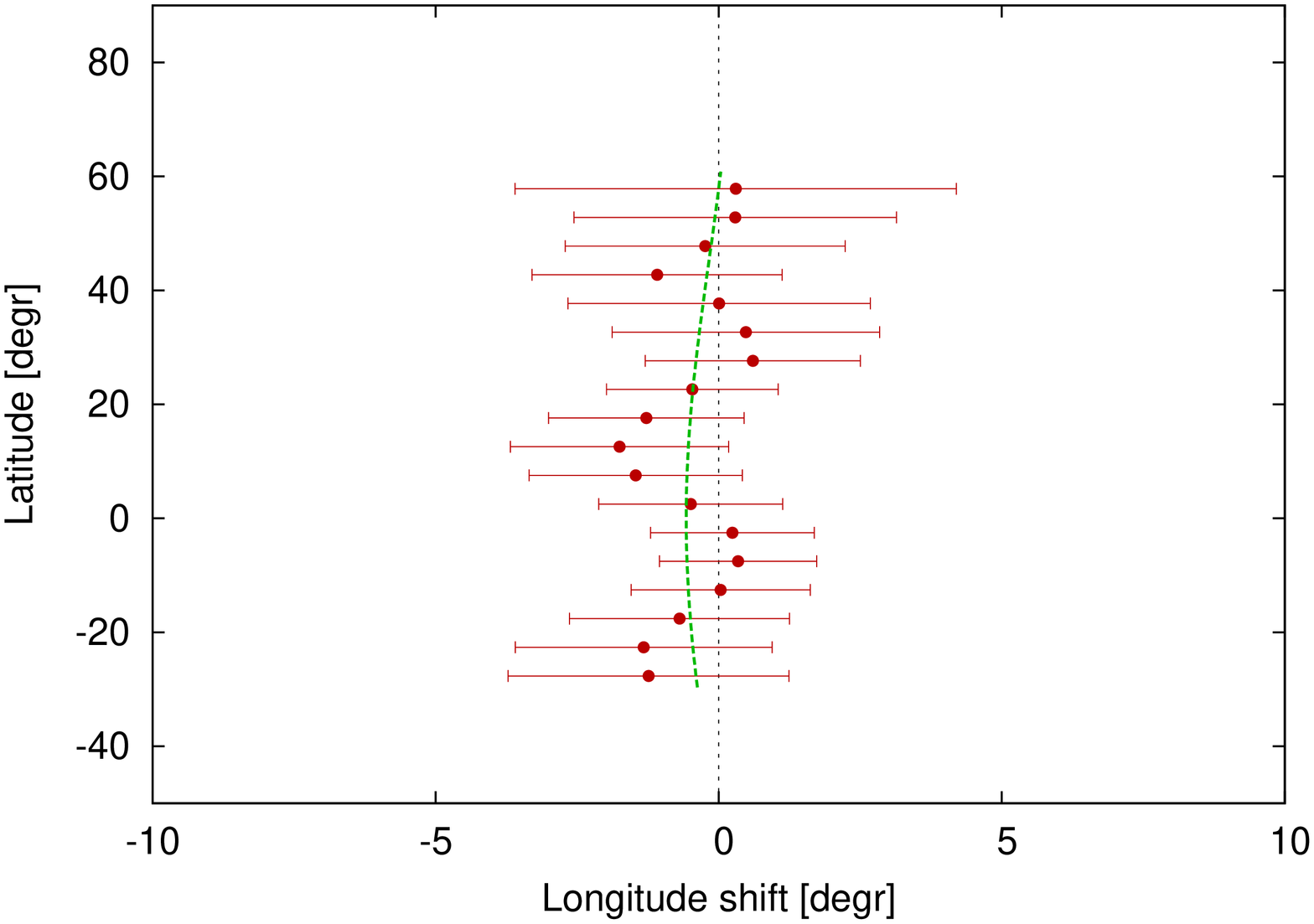}

 \includegraphics[angle=-90,width=0.97\columnwidth]{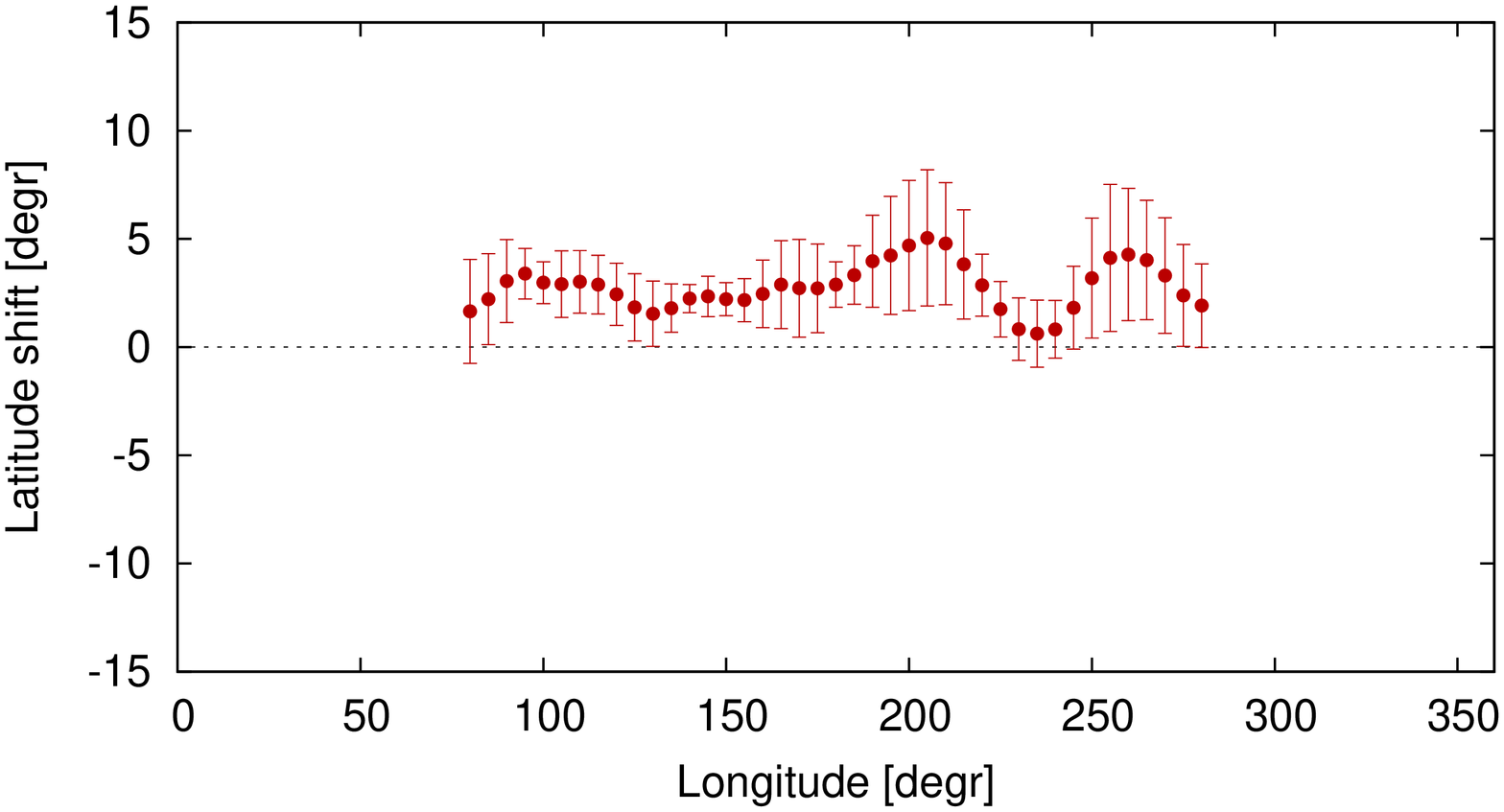}
\caption{Zonal (top) and meridional (bottom) flow components from
the LCT method. The zonal distribution may be indicative of anti-solar
differential rotation with a shear of $\alpha=-0.0022\pm0.0016$.
The meridional component shows a coherent poleward flow with an
average velocity of $2.5\pm0.1$\degr\ over a rotation cycle.}
\label{zonalmerid}
\end{figure}

In Fig.~\ref{zonalmerid} we plot the zonal and
meridional flow components extracted from the LCT map. Despite its
large error bars,
the latitudinal distribution of the zonal component is in favor of a weak
anti-solar differential rotation (DR afterwards). The latitude ($\beta$) dependent rotation
law is assumed in the usual quadratic form of
\begin{equation}\label{drlaw}
\Omega (\beta) = \Omega_{\rm eq}  - \Delta\Omega \sin^2 \beta \ ,
\end{equation}
where $\Omega_{\rm eq}$ is the
equatorial angular velocity, $\Delta\Omega= \Omega_{\rm eq} - \Omega_{\rm pole}$
is the angular velocity difference between the equator and the pole. The
surface shear parameter is defined as $\alpha = \Delta\Omega /
\Omega_{\rm eq} $. The best fit gives $\Omega_{\rm eq}=18.33\pm0.01$\degr/day
and $\Delta\Omega=-0.04\pm0.03$\degr/day which yields
$\alpha=-0.0022\pm0.0016$, i.e., an almost rigid body rotation.
The meridional flow component indicates mostly poleward spot migration.
The formal surface-averaged poleward meridional component is $2.5\pm0.1$\degr\
over a rotation cycle of $\approx$20 days, which converts to an average velocity of
$\approx220\pm10$\msec.
In the RS\,CVn-type $\sigma$\,Gem such a meridional flow could be maintained by
large-scale thermal inhomogeneities (from large cool spots) and/or by tidal effects
from the close binary nature (cf. Kitchatinov \& R\"udiger \cite{kitrued}).

\section{Cross-correlation results}
\label{accord}

In this section we compare the LCT results with our recent findings
(K\H{o}v\'ari et al. \cite{AARNsgem})
for the same data by
applying the method of `Average cross-correlation of contiguous Doppler images'
(hereafter ACCORD). The method was described and applied to detect surface
differential rotation first for LQ\,Hya in K\H{o}v\'ari et al. (\cite{AAlqhya}),
and further on has become a powerful tool in analysing time-series
Doppler maps (e.g., K\H{o}v\'ari, Weber \& Strassmeier \cite{koweb},
K\H{o}v\'ari et al. \cite{AAzetand}).

We use again the consecutive but contiguous correlation image pairs
(i.e., \#1$-$\#3, \#2$-$\#4, \#3$-$\#5 and \#4$-$\#6) and compute four
cross-correlation function (ccf) maps. Then we make a linear
normalization, since the time baseline of the ccf maps are different.
After averaging the normalized maps,
we fit the correlation peak for each latitudinal stripe with a
Gaussian profile. These Gaussian peaks per latitude are fitted with
the usual quadratic form (Eq.~1).
The result shown in the upper panel of
Fig.~3 suggests an anti-solar type DR law with
$\Omega_{\rm eq}=18.29 \pm 0.05$\degr/day and $\Delta\Omega=-0.38 \pm 0.08$\degr/day
and a surface shear of $\alpha\approx-0.021\pm0.005$.

Meridional motion of surface features can be derived similarly,
by cross-correlating the
corresponding longitude stripes along the meridian circles
(but restricting only to the more reliable visible hemisphere).
The resulting average latitudinal ccf map is plotted in the lower panel of Fig.~3.
The best correlating latitudinal shifts suggest a common poleward migration
of spots of $\approx4.1\pm0.3$\degr\ per rotation cycle, which can be
interpreted as a poleward meridional flow with an average
velocity of $\approx 350$\msec.

\begin{figure}[t]
\includegraphics[width=1.0\columnwidth]{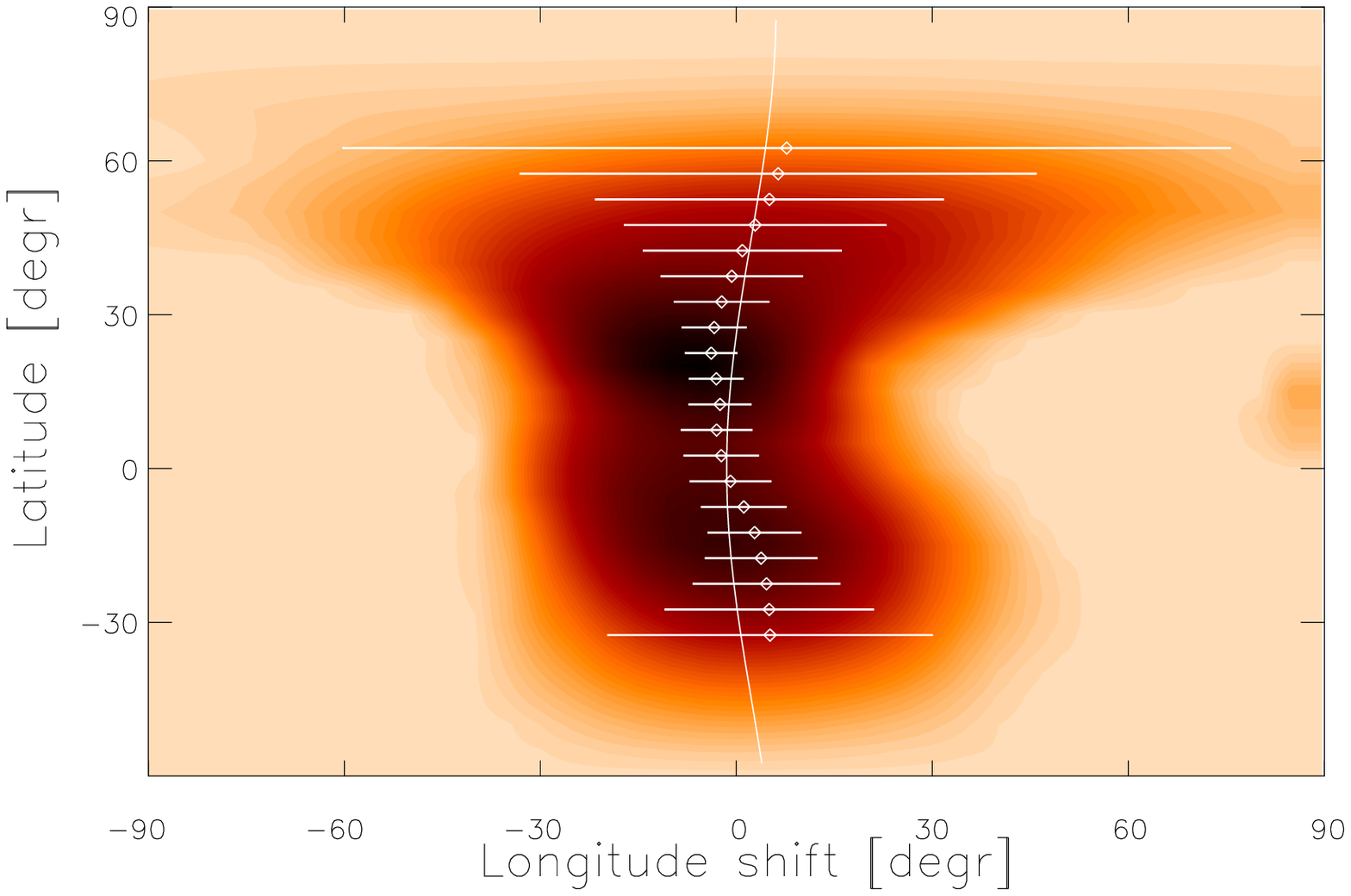}

\includegraphics[angle=0,width=1.0\columnwidth,height=0.5\columnwidth]{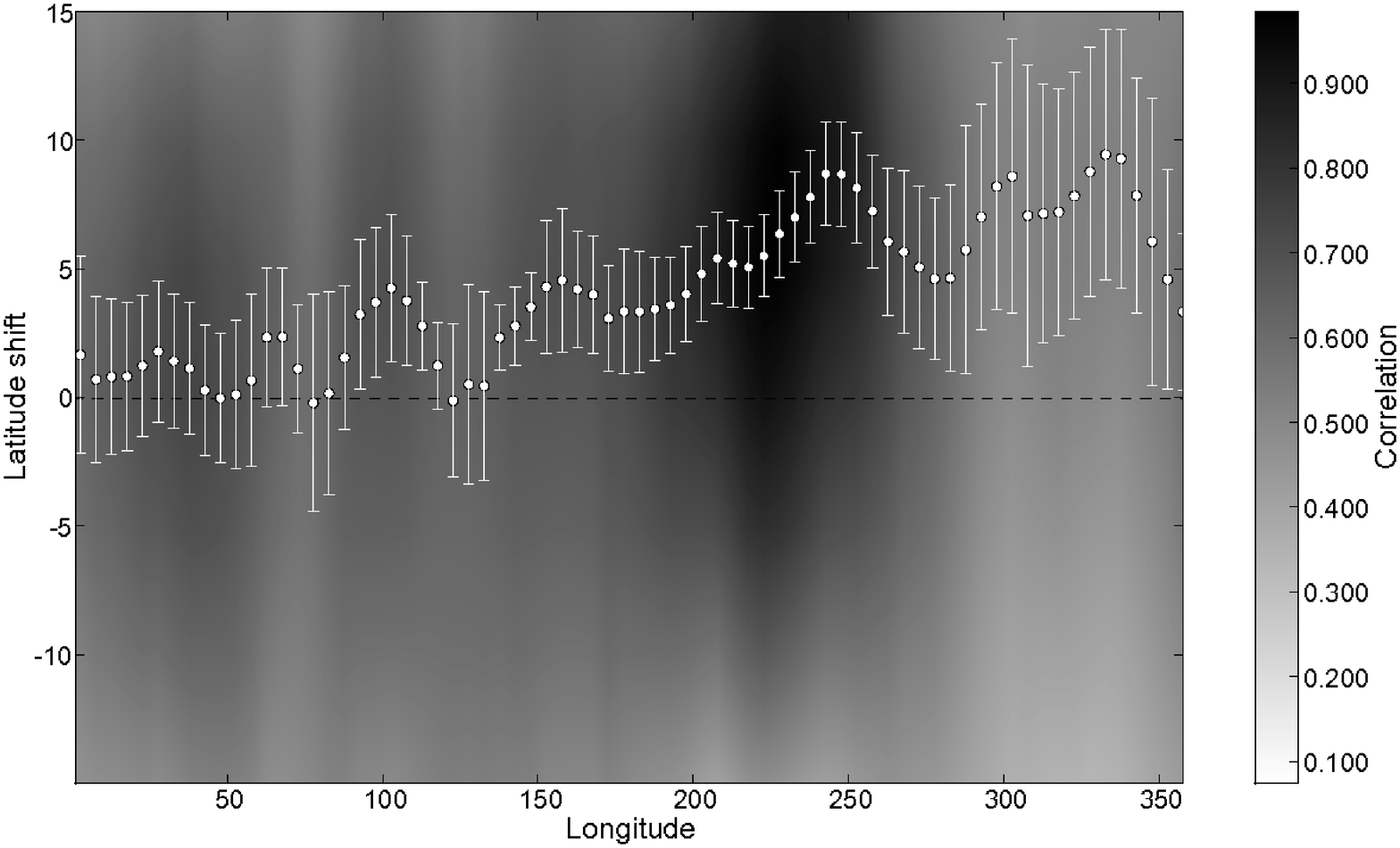}
\caption{Latitudinal (top) and longitudinal (bottom) cross-correlation maps
of $\sigma$\,Gem from ACCORD using the average (Ca+Fe) Doppler images
from K\H{o}v\'ari et al. (\cite{AAsgem}, \cite{AARNsgem}),
suggesting anti-solar DR with a coherent poleward migration of spots.}
\label{ccfs}
\end{figure}

\section{Summary and conclusions}\label{disc}

We have demonstrated that LCT technique can be used to map flow
fields on stellar surfaces. Applying LCT for time-series Doppler images
of $\sigma$~Gem we reconstruct large-scale surface
flows in the order of several hundreds of \msec\ and observe complex network of convergent flows around spots.
Zonal flow components suggest rigid-body rotation or anti-solar DR with a very weak surface shear
of $\alpha=-0.0022\pm0.0016$,
while meridional flow components show a coherent poleward spot migration
at an average velocity of $220\pm10$\msec. For a more detailed description of
the method as well as for numerical tests on artificial data, we refer to
a forthcoming paper (\v{S}vanda et al. \cite{lctlq}).

Our findings agree well with the results from a recent application of the ACCORD
method on the same data (K\H{o}v\'ari et al. \cite{AARNsgem}).
ACCORD resulted also in anti-solar DR law, but with
a stronger shear of $\alpha=-0.021\pm0.005$. From the latitudinal
cross-correlation a coherent poleward flow was derived
with a rate of $\approx350$\msec, which is in qualitative and also in quantitative agreement
with the meridional flow component from LCT.
The deviation could be originated from the different limitations
of the two different approaches.

Sufficiently fast meridional flow can result in anti-solar type DR and the
necessary angular momentum transport (i.e., the necessary
rate of the meridional flow)
can be estimated from Eq.~15 in Kitchatinov \& R\"udiger (\cite{kitrued}).
Taking a turnover time of $\tau=5.5\times10^6$\,s with a
mixing length $l$ of $7\times10^8$\,m (cf. Gunn, Mitrou \& Doyle \cite{gumido98},
Patern\`o et al. \cite{pat02}) would yield $\approx300$\msec, which agrees with
our results derived from two different methods. If the poleward
surface migration on $\sigma$\,Gem is interpreted due to an underlying
meridional circulation, our result could be regarded as a confirmation of
the theoretical estimates in Kitchatinov \& R\"udiger (\cite{kitrued}).

\begin{acknowledgements}
ZsK, KV, KO and EF-D are supported by the Hungarian Science
Research Program (OTKA) grant T-048961. ZsK
is a grantee of the Bolyai J\'anos Scholarship of the HAS. M\v{S} is supported by the Czech Science
Foundation under grant 205/03/H144. The Astronomical Institute of
Academy of Sciences of the Czech Republic is working on the Research
project AV0Z10030501. KGS thanks the U.S. National Solar Observatory
for the possibility to record long time series of
stellar data during the late night-time program at the
McMath-Pierce telescope.
\end{acknowledgements}



\begin{thebibliography}{}

\bibitem[2001]{amb}
Ambro\v{z}, P.: 2001. Sol.~Phys. 198, 253

\bibitem[1998]{gumido98}
Gunn, A.~G., Mitrou, C.~K., Doyle, J.~G.: 1998, MNRAS 296, 150

\bibitem[2004]{kitrued}
Kitchatinov, L.L., R\"udiger, G.: 2004, AN 325, 496

\bibitem[2001]{AAsgem}
K\H{o}v\'ari, Zs., Strassmeier, K.~G., Bartus, J., Washuettl, A.,
Weber, M., Rice, J.~B.: 2001, A\&A 373, 199

\bibitem[2004]{AAlqhya}
K\H{o}v\'ari, Zs., Strassmeier, K.~G., Granzer, T., Weber, M.,
Ol\'ah, K., Rice, J.~B.: 2004, A\&A 417, 1047

\bibitem[2005]{koweb}
K\H{o}v\'ari, Zs., Weber, M., Strassmeier, K.~G.: 2005, in: Cool Stars,
Stellar Systems and the Sun 13, 5-9 July 2004, Hamburg, Germany, ESA-SP-560 (Vol. II), 731

\bibitem[2007a]{AAzetand}
K\H{o}v\'ari, Zs., Bartus, J., Strassmeier, K.~G., Ol\'ah, K.,
Weber, M., Rice, J.B., Washuettl, A.: 2007a, A\&A 463, 1071

\bibitem[2007b]{AARNsgem}
K\H{o}v\'ari, Zs., Bartus, J., Strassmeier, K.~G., Vida, K., \v{S}vanda, M.,
Ol\'ah, K.: 2007b, A\&A 474, 165

\bibitem[2002]{pat02}
Patern\`o, L., Belvedere, G., Kuzanyan, K.~M., Lanza, A.~F.: 2002, MNRAS 336, 291

\bibitem[1999]{sobo99}
Sobotka, M., V\'azquez, M., Bonet, J.~A., Hanslmeier, A., Hirzberger, J.: 1999, ApJ 511, 436

\bibitem[2000]{sobo00}
Sobotka, M., V\'azquez, M., Cuberes, M.~S., Bonet, J.~A., Hanslmeier, A.: 2000, ApJ 544, 1155

\bibitem[2006]{sva2}
\v{S}vanda, M., Klva\v{n}a, M., Sobotka, M.: 2006, A\&A 458, 301

\bibitem[2008]{lctlq}
\v{S}vanda, M., K\H{o}v\'ari, Zs., Strassmeier, K.~G.: 2008, A\&A,
in prep.


\end{thebibliography}
\end{document}